\documentclass[twocolumn,showpacs,preprintnumbers,amsmath,amssymb,prl]{revtex4}
\usepackage{graphicx}
\usepackage{dcolumn}
\usepackage{bm}
\begin{document}

\title{Origin of the high-energy kink or the waterfall effect in the
photoemission spectrum of the ${\rm Bi_2Sr_2CaCu_2O_8}$ high-temperature
superconductor}
\author{Susmita Basak$^1$, Tanmoy Das$^1$, Hsin Lin$^1$, J.
Nieminen$^{1,2}$, M. Lindroos$^{1,2}$, R.S. Markiewicz$^{1,3,4}$, and A.
Bansil$^1$}
\address{$^1$ Physics Department, Northeastern University, Boston MA
02115, USA\\
$^2$ Institute of Physics, Tampere University of Technology, P.O. Box
692, 33101 Tampere, Finland\\
$^{3}$\mbox{SMC-INFM-CNR, Dipartimento di Fisica, Universit\`a di Roma
 ``La Sapienza'', P. Aldo Moro 2, 00185 Roma, Italy.}\\
$^{4}$\mbox{ISC-CNR, Via dei Taurini 19, 00185 Roma }\\
}

\date{\today}
\begin{abstract}
The high-energy kink or the waterfall effect seen in the photoemission
spectra of the cuprates is suggestive of the coupling of the
quasiparticles to a high energy bosonic mode with implications for the
mechanism of superconductivity. Recent experiments however indicate
that
this effect may be an artifact produced entirely by the matrix element
effects, i.e. by the way the photoemitted electron couples to the
incident
photons in the emission process. In order to address this issue
directly,
we have carried out realistic computations of the photo-intensity in
${\rm Bi_2Sr_2CaCu_2O_8}$ (Bi2212) where the effects of the matrix
element are included together with
those of the corrections to the self-energy resulting from electronic
excitations. Our results demonstrate that while the photoemission
matrix element plays an important role in shaping the spectra, the
waterfall
effect is a clear signature of the presence of strong coupling of
quasiparticles to electronic excitations.
\end{abstract}
 \maketitle\narrowtext

An anomalous `high-energy kink' (HEK) in the dispersion, which gives
the associated angle-resolved photoemission spectrum (ARPES) the
appearance of a `waterfall' was first seen\cite{graf} in ${\rm
Bi_2Sr_2CaCu_2O_8}$ (Bi2212) cuprate superconductor. Such HEKs or
`waterfall' effects have now
been established as being a universal feature in the cuprates, and are
found along both nodal and antinodal directions with their energy and
momentum scales varying with doping and details of the band
structure\cite{graf,moritz}. The HEK has been interpreted theoretically
as providing evidence for interaction of the quasiparticles with some
bosonic mode of the system\cite{Mwater,MJMS} with implications for the
mechanism
of high-temperature superconductivity. Recent experiments show however
that the HEK is quite sensitive to matrix element (ME) effects, i.e. to
the nature of the potoemission process or the way the incident photon
couples with the electronic states of the system in generating the
photoemitted electrons. In particular, the ARPES spectra undergo
substantial changes in shape as one probes the electronic states by
varying the energy of the incident photon or the momentum of the outgoing
electron.\cite{Int,Zhou,DanD} In Bi2212, for example, the shape of the
ARPES spectrum varies more or less periodically with photon energy from a
shape that displays a single band-tail with relatively large intensity
giving the spectrum a `Y' shape, to a spectrum which
shows the presence of double-tails with a waterfall.\cite{Int2}
These results have led to speculation that the
HEK may be an artifact produced entirely by ME effects.\cite{Int,Int2}
On the other hand, for all polarizations and photon energies, the near
Fermi surface features are strongly renormalized from LDA values, whereas
the band bottom is not.  Such a change in dispersion with energy is
generally taken to be a genuine signature of a bosonic mode coupling.

In order to address this controversy, we have carried out extensive first
principles, one-step photo-intensity computations in Bi2212 in which we
include not only the effects of the ARPES matrix element, but also
incorporate a model self-energy based on accurate susceptibility
calculations which properly reproduce the HEK
phenomenology\cite{Mwater,tanmoy6,LDA}.  In this way, we establish
conclusively that despite a strong modulation of the spectra due to the
ARPES matrix element, a genuine HEK or a waterfall effect is still
present in the cuprates, and that its presence indicates a significant
coupling to bosons of electronic origin. Given the strength of the
coupling and the high associated energy scale, these bosons are likely to
play an important role in both the Mott and the superconducting physics
of the cuprates.  In this connection, we also discuss model ARPES
computations based on simplified tight-binding models for the purpose of
gaining a handle on the interplay between the matrix element and
self-energy effects, and also for delineating the nature of the striking
characteristics of the ARPES matrix element such as the crossover from
the Y-type spectral shape to the waterfall shape with photon energy.

\begin{figure}[htp]
\centering
\rotatebox{0}{\scalebox{0.35}{\includegraphics{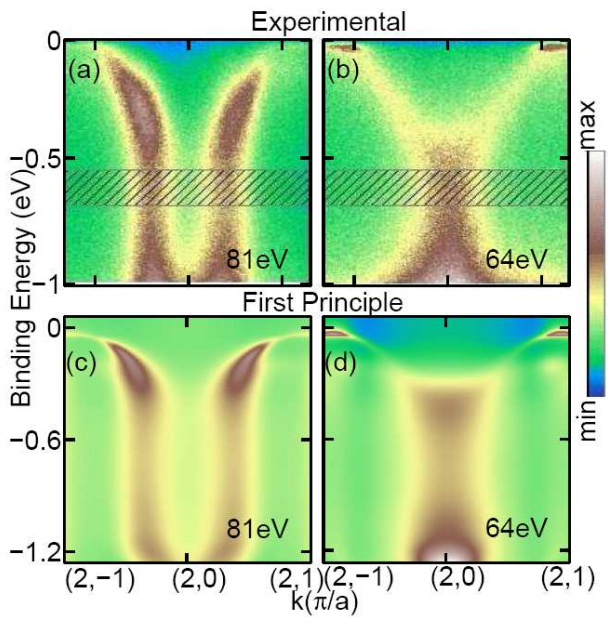}}}
\caption{(Color online.)
ARPES spectra in Bi2212.
(a) and (b) are experimental results in the second Brillouin zone
at photon energies of 81 eV and 64 eV, respectively.\cite{Int2}
(c) and (d) are corresponding theoretical photoemission
spectra based on first principles, one-step calculations.
}
\label{Bi}
\end{figure}

We discuss our key findings with reference to Fig. 1. Note first how the
shape of the experimental spectrum changes dramatically at different
photon energies. In panel (a) at 81 eV, the spectral intensity presents
the appearance of a pair of `waterfalls' with a region of low intensity
through the middle of the figure. This is in sharp contrast to the
measured spectrum in (b) at 64 eV where we see a `Y-shape' with the two
arms of the Y connecting a vertical region of high intensity. Our
realistic first-principles photo-intensity computations in which the
matrix element as well as self-energy effects are accounted for reproduce
the characteristic features of these shapes: The waterfall shape in panel
(c) at 81 eV and the Y-shape in panel (d) at 64 eV, even though the arms
of the Y in the computed spectrum in (d) are less clearly defined than
the corresponding measurements in (b).

For the purpose of delineating the roles of the self-energy and matrix
elements in shaping the ARPES spectra, Fig.~2 presents
the results of photo-intensity computations based
on a simplified two-band tight-binding Hamiltonian which models the
low-energy electronic structure of Bi2212.\cite{Arun3,footAll} The real and imaginary
parts of the self-energy\cite{footAll} within the framework of the
two-band model
are shown in panel (a) for the antibonding band (AB) [similar results for
the bonding band (BB) are not shown for brevity].  The model bands
dressed by only the real part of the self-energy are shown in
(b), together with the LDA bands at $k_z=2\pi/c$ (magenta lines). The
real part of the self-energy [solid blue line in (a)] is almost linear in
$\omega$ in the low-energy region. We write the slope of the linear
part as $(1-Z^{-1})$ to define the renormalization coefficient $Z$.  This
leads to a renormalized quasiparticle dispersion $\bar{\xi_{\bf
k}}=Z\xi_{\bf k}$, which, e.g., reduces the bi-layer splitting between
the AB and BB bands at the antinodal point to $Z(\xi_{\pi,0}^{AB}-
\xi_{\pi,0}^{BB})$, consistent with experimental results\cite{footAll,Int}.
In contrast, at the $\Gamma-$point the dressed bands determined by
$\xi_{\Gamma}^{AB/BB}+ \Sigma^{\prime}(\xi_{\Gamma}^{AB/BB})$ move
further away from each other.
This is due to the fact that in Bi2212, $\xi_{\Gamma}^{AB/BB}$ lies in an
energy range where the corresponding self-energy values
$\Sigma^{\prime}(\xi_{\Gamma}^{AB/BB})$ possess an opposite sign.
Moreover, the Green's function for the anti-bonding band shows multiple
poles at the $\Gamma-$point because the line $\omega-\xi_{\Gamma}^{AB}$
coincides more than once with $\Sigma^{\prime}(\omega)$ in (a), and as
a result the anti-bonding band develops an additional splitting. These
opposing tendencies at low and high energies, which are seen clearly in
the experimental spectra, are an unambiguous signature of strong
coupling to a bosonic mode at intermediate energy.

\begin{figure}[htp]
\centering
\rotatebox{270}{\scalebox{0.4}{\includegraphics{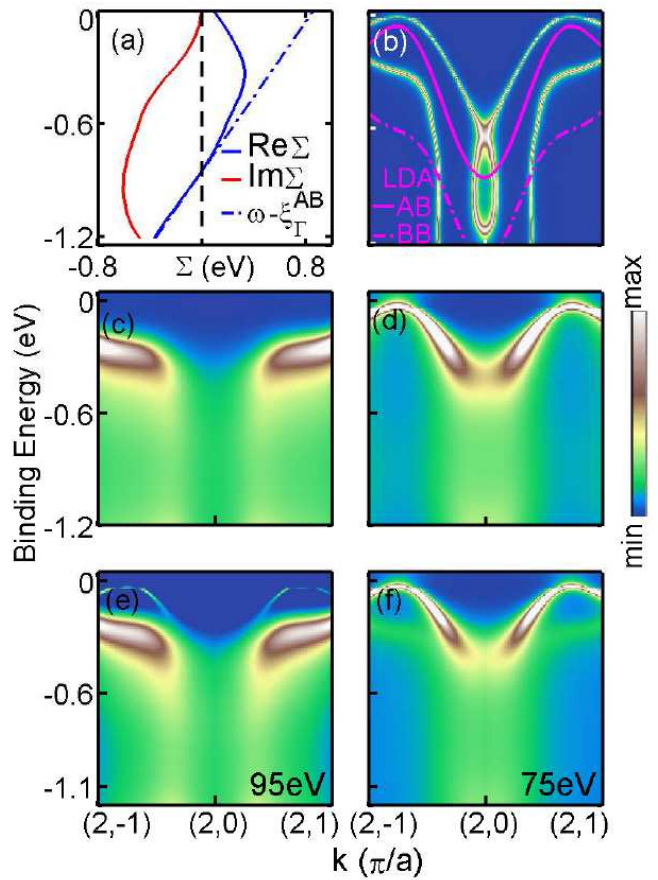}}}
\caption{(Color online.)
Model self-energy and spectral weight in Bi2212.
(a) Real (blue solid line) and imaginary (red solid line) parts of the
computed self-energy used in photo-intensity computations. The blue dashed
line
gives $\omega-\xi^{AB}_{\Gamma}$. (b) Dispersion renormalized by the real part
of the self-energy in (a) is compared to the bare dispersion of the
anti-bonding
band (AB, magenta solid line) and the bonding band (BB, magenta dashed
line).  (c) and (d) Spectral weights dressed by real and imaginary part
of the self-energy for AB and BB, respectively.
(e) and (f) Final photoemission intensities obtained
after incorporating the matrix elements at two different
photon energies, $95$ eV and $75$ eV, respectively.
}
\label{Bi}
\end{figure}

When the imaginary part of the self-energy $\Sigma^{\prime\prime}$ is
turned on, interesting spectral weight modulations emerge, which are
shown separately in Fig.~2(c) and 2(d) for the AB and BB bands,
respectively.  $\Sigma^{\prime\prime}$ plays a crucial role in
redistributing spectral weight such that the weight is shifted from the
coherent region near the Fermi energy into incoherent parts at higher
energies to produce the HEK features seen experimentally\cite{Mwater}.
However, the differences noted earlier near $\Gamma$ lead to striking
differences in the manifestation of the kink effect in the AB and BB
spectra.  In AB, the band bottom at $\Gamma$ lies above the `waterfall'
region, and $\Sigma^{\prime\prime}$
creates a long tail in the dispersion extending to high
energies,\cite{foot4} so the fully dressed band exhibits the `Y'-shaped
pattern of panel (c).  In contrast, the `waterfall' shape emerges
distinctly in the bonding band in panel (d) where the band bottom lies
below the boson peak. These two dispersions are remarkably close to the
two experimental dispersions observed at different photon energies as
seen in panels 1(a) and 1(b)\cite{Int2} and provide insight into the
nature of the waterfall phenomenon in the spectra.

In order to delineate contributions from different orbitals, we have
computed the photo-intensity within the framework of our tight-binding
model using the linear electron-photon interaction. The matrix element
for the $\mu^{th}$ band is then shown straightforwardly to be given
by\cite{foot7}
\begin{eqnarray}\label{eq:1}
M_\mu({\bf k}_f)&=&\hbar A{\hat{\epsilon}} \cdot {{\bf
k}_f}\sum_{i,nlm}(-i)^lY_{lm}(\theta_{{k}_f},\phi_{{k}_f})F_{nl}(k_f)
\nonumber \\
&\times&\langle i,lmn|\mu \rangle e^{-i{{\bf k}_f}\cdot{\bf{R}}_i}
\end{eqnarray}
Here ${\bf k}_f$ is the momentum of the ejected electron,
$\hat\epsilon$
denotes the polarization of light with vector potential $A$, and
$Y_{lm}$
is the spherical harmonic for the angular variables of ${\bf k}_f$. The
final state is taken to be a free electron state. The initial state
$|\mu\rangle$ is a tight-binding state, which is expanded into atomic
orbitals $(nlm)$ of the $i^{th}$ atom in the unit cell at position
$\bf {R_i}$. The form factor

\begin{eqnarray}\label{eq:3}
F_{nl}(k_f)=\int r^2drj_{l}(k_fr)R_{nl}(r),
\end{eqnarray}
where $j_l$ is a spherical Bessel function, is evaluated numerically
using
the radial part of the atomic wavefunction\cite{foot5,hermanson,Atomic}.
Fig. 3 shows that at low $k_f$ the oxygen contribution $F_{21}$ for the
oxygen $p_x$ and $p_y$ orbitals is dominant, while the Cu contribution
$F_{32}$ for
the Cu $d_{x^2-y^2}$ orbital dominates for $k_f>2~a.u.^{-1}$.
 Although Eq. 2 is general, we have used only
three orbitals, i.e. Cu $d_{x^2-y^2}$, O $p_x$, and O $p_y$ in
expanding
the AB and BB bands to obtain the photo-intensities for our
illustrative
purposes. Finally, Eq. 1 can be recast into a useful form by
collapsing all the symmetry information concerning the $i^{th}$ orbital
into the structure factor $S_i^{\mu}({\bf k}_f)$:
\begin{eqnarray}\label{eq:2}
M_\mu({\bf k}_f)= \sum_i S_i^{\mu} ({\bf
k}_f)e^{-i{\bf{k}_f}\cdot{\bf{R}}_i}
\end{eqnarray}

\begin{figure}[htp]
\centering 
\rotatebox{0}{\scalebox{0.3}{\includegraphics{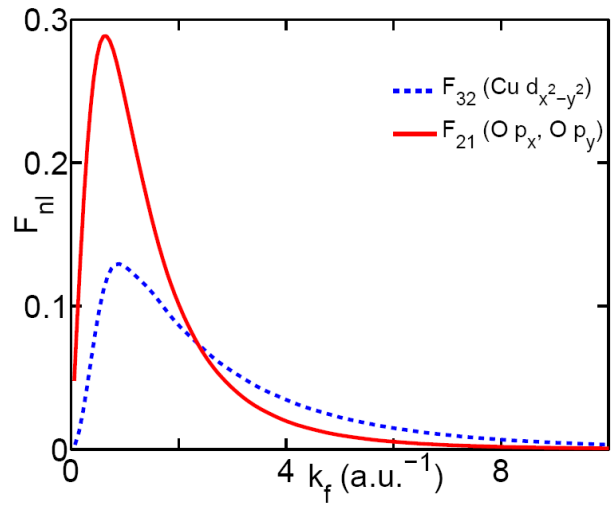}}}
\caption{Form factors of Eq. 1 for copper $d_{x^2-y^2}$ orbital (solid
line), $F_{32}$, and for oxygen p orbitals, $F_{21}$.
} \label{Fnl}
\end{figure}

For a bi-layer system, the structure factor of Eq. 3 becomes
\begin{equation}
S^{\pm}_{\bf G}=S^0_{\bf G}(h,k)[1\pm e^{-ik_f^\perp d}], \label{eq:6}
\end{equation}
where $S^0$ is the two-dimensional structure factor of Eq.~3, the +
sign refers to the AB band and the - sign to the BB band.
${\bf G}=2\pi (h/a,k/a,\ell /c)$ is a reciprocal lattice vector
with $a$ being the in-plane and $c$ the out-of-plane lattice constant,
$d$ denotes the separation of the CuO$_2$ layers in a
bilayer.\cite{foot6,dimpling}
The phase factor in Eq. 4 depends on the photon energy through
\cite{Int2,Feng2002}
\begin{equation}
k_f^{\perp}=\sqrt{{2m\over\hbar^2}(h\nu-E_{bind}-\Phi+V_0)-(k_{\parallel}+
n_{\parallel}G_{\parallel})^2},
\label{eq:5}
\end{equation}
where $h\nu$ is the incident photon energy and $E_{bind} \approx 0.6$ eV
is
the binding energy of the electron in the solid at the waterfall, $\Phi
\approx 4$ eV is the work function, $V_0 = 10$ eV is the inner
potential of the crystal and $k_{\parallel}+n_{\parallel}
G_{\parallel}$ is the total in-plane wave number, which we have taken to
be
$\approx 2\pi/a$ to match the experimental conditions.\cite{foot8} Thus, whenever
$k_f^{\perp}d$ changes by $\pi$, the spectrum would switch from the odd
to
the even bilayer, a change that can be induced in view of Eq. 5 via
the
photon frequency $\nu$. This behavior is indeed seen in panels (e) and
(f)
of Fig. 2 where the matrix element is incorporated in the
photo-intensity
computations using Eqs. 1-5. In particular, at $75$ eV in panel (f), the
AB band gets highlighted resulting in a Y-shaped spectrum with a tail
extending to high energies. In contrast, in panel (e) at $95$ eV, the
bonding band dominates and spectral shape reverts to that of a waterfall
with a double tail.

\begin{figure}[htp]
\centering \hskip-.25 cm
\rotatebox{270}{\scalebox{0.42}{\includegraphics{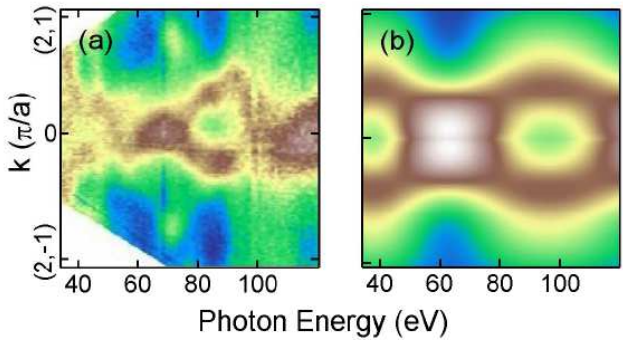}}}
\caption{(Color online)
Spectral weight integrated over the shaded binding energy window of
Fig. 1(a) or 1(b) in the intermediate energy region is shown to highlight
how the spectra vary periodically between the Y- and waterfall shapes as a
function of photon energy. (a) Experimental spectral weight normalized
to the peak intensity at each energy.~\protect\cite{Int2}; (b) Theoretical
weights for the bilayer split spectra of Figs.~1(e) or (f). The colorbar
is the same as Fig.~1.}
\label{Bi}
\end{figure}
%

Along the preceding lines, Fig. 4 further discusses the Y-waterfall
spectral shape oscillation as a function of the
photon energy. For this purpose, we consider in Fig. 4 the integrated
spectral weight over the shaded binding energy window shown in Fig.
1(a) or 1(b). The Y-shape is then characterized by a relatively narrow
single tail in momentum (vertical axis in Fig. 4), while the waterfall
displays a
splitting of this feature due to the presence of two tails. The
experimental results of Ref.~\onlinecite{Int2} shown in Fig. 4(a) are
seen
to be in good accord with the corresponding computations in Fig. 4(b),
which are based on the bilayer dispersion of Ref.~\onlinecite{kzdisp},
some differences between theory and experiment with respect to the
onset
of Y- or waterfall shape in photon energy notwithstanding. Notably, the
theoretical results of Fig.~4(b) account for the effects of `dimpling'
of
the CuO$_2$ planes (i. e., deviations of the positions of the Cu and O
atoms
in the crystal structure of Bi2212 from a perfectly planar geometry) by
including two bilayer splittings $d_1$ and $d_2$ in modeling the
bilayer.
When the self-energy correction is then included in the binding energy,
$E_{bind}=\xi_{k_{\parallel}} +\Sigma(\xi_{k_{\parallel}}) $, it makes
$k_f^{\perp}$ complex in Eq. 5. This complex $k_f^{\perp}$ damps the
contribution from the lower CuO$_2$ plane and thus modifies the
intensity
pattern through interference effects. We find that these corrections
do not make much difference in our result. Taken as a
whole, the comparisons between the first-principles and model
computations
of the photo-intensity in Figs. 1, 2 and 4 also indicate that our simpler
model with a limited number of orbitals is capable of capturing many
salient spectral features in the low energy region and giving insight
into
their origin.


In summary, we have carried out extensive computations of the
photo-intensity in Bi2212 where the effects of the photoemission matrix
element as well those of the coupling of the quasiparticles to
electronic excitations are included realistically. We thus establish that
despite
the importance of the matrix element in shaping the spectra, the
waterfall
effect is a clear signature of the coupling of the electronic system to
a high-energy bosonic mode, which bears on the physics of the pseudogap
and the mechanism of high-temperature superconductivity. Our analysis of
the spectra based on a simplified two-band tight-binding model reveals
how
the near-Fermi energy bonding and anti-bonding bands associated with the
CuO$_2$ bilayers in Bi2212 produce characteristic Y-shape and waterfall
shape
of the spectrum as a function of the energy of the incident photons. Such
a periodic modulation of the spectrum with photon energy may provide a
new spectroscopic tool for getting a handle on the structural aspects of
the bilayer via the photoemission technique.

\begin{acknowledgments}
We thank J. Lorenzana for useful suggestions.
This work is supported by the US Department of Energy,
Office of Science, Basic Energy Sciences contract DE-FG02-
07ER46352, and benefited from the allocation of
supercomputer time at NERSC, Northeastern University's
Advanced Scientific Computation Center (ASCC), and the
Institute of Advanced Computing, Tampere.
RSM's work has been partially funded by the Marie Curie
Grant PIIF-GA-2008-220790 SOQCS.
\end{acknowledgments}

%


\begin{thebibliography}{99}
\bibitem{footAll} A $k_z$-dispersion is also included for the initial
state with a TB parameter $t_z=0.076$ eV.
Self-energy is evaluated as in
Ref.~\onlinecite{tanmoy6},
except that the effect of bilayer splitting is included.  This turns out
to be a minor correction.  Doping is fixed to the optimal regime of
Bi2212 with Hubbard $U= 1$~eV to obtain a self-consistent
renormalization value of $Z=0.5$.
Our bare band bilayer splitting varies from
around $120$ meV at $\Gamma$ to $\sim 340$ meV at $(\pi,0)$; the
corresponding dressed values are $\sim 60$ meV
and $150$ meV, respectively.
\bibitem{foot4}We find that tails are typically produced at band
bottoms due to the incoherent spectral weight introduced by the self
energy.
\bibitem{foot5}An additional ME effect due to final state coupling to the
detector \cite{hermanson} does not significantly alter these results,
and is therefore not included.
\bibitem{foot6}We take the dimpling of the CuO$_2$ planes into account
by using two different bilayer distances; $d_1=2.9 \AA$ for oxygen and
$d_2=3.2 \AA$ for copper \cite{dimpling}.
\bibitem{foot8} The value of the inner potential was obtained from an LDA
Slab calculation.
\end{thebibliography}
\end{document}